\documentclass[twocolumn]{aastex63}

\newcommand{\hawki}{\mbox{HAWK-I}}
\shorttitle{Deep near-infrared photometric maps of Local Group Galaxies}
\shortauthors{Karczmarek et al.}

\graphicspath{{./}{figures/}}

\begin{document}

\title{The Araucaria Project: Deep near-infrared photometric maps of Local and Sculptor Group galaxies.\\I. Carina, Fornax, Sculptor}

\correspondingauthor{Paulina Karczmarek}
\email{pkarczmarek@astro-udec.cl}

\author[0000-0002-0136-0046]{Paulina Karczmarek}
\affiliation{Universidad de Concepci\'{o}n, Departamento de Astronom\'{i}a, Casilla 160-C, Concepci\'{o}n, Chile}
\affiliation{Nicolaus Copernicus Astronomical Center, Polish Academy of Sciences, Bartycka 18, 00-716, Warsaw, Poland}

\author{Grzegorz Pietrzy\'{n}ski}
\affiliation{Nicolaus Copernicus Astronomical Center, Polish Academy of Sciences, Bartycka 18, 00-716, Warsaw, Poland}

\author[0000-0003-1405-9954]{Wolfgang Gieren}
\affiliation{Universidad de Concepci\'{o}n, Departamento de Astronom\'{i}a, Casilla 160-C, Concepci\'{o}n, Chile}

\author{Weronika Narloch}
\affiliation{Universidad de Concepci\'{o}n, Departamento de Astronom\'{i}a, Casilla 160-C, Concepci\'{o}n, Chile}

\author{Gergely Hajdu}
\affiliation{Nicolaus Copernicus Astronomical Center, Polish Academy of Sciences, Bartycka 18, 00-716, Warsaw, Poland}

\author{Gonzalo Rojas Garc\'{i}a}
\affiliation{Nicolaus Copernicus Astronomical Center, Polish Academy of Sciences, Bartycka 18, 00-716, Warsaw, Poland}

\author{Miko{\l}aj Ka{\l}uszy\'{n}ski}
\affiliation{Nicolaus Copernicus Astronomical Center, Polish Academy of Sciences, Bartycka 18, 00-716, Warsaw, Poland}

\author[0000-0002-3125-9088]{Marek G\'{o}rski}
\affiliation{Nicolaus Copernicus Astronomical Center, Polish Academy of Sciences, Bartycka 18, 00-716, Warsaw, Poland}

\author[0000-0001-6118-5844]{Ksenia Suchomska}
\affiliation{Nicolaus Copernicus Astronomical Center, Polish Academy of Sciences, Bartycka 18, 00-716, Warsaw, Poland}

\author[0000-0002-7355-9775]{Dariusz Graczyk}
\affiliation{Nicolaus Copernicus Astronomical Center, Polish Academy of Sciences, Rabia\'{n}ska 8, 87-100, Toru\'{n}, Poland}

\author[0000-0003-3861-8124]{Bogumi{\l} Pilecki}
\affiliation{Nicolaus Copernicus Astronomical Center, Polish Academy of Sciences, Bartycka 18, 00-716, Warsaw, Poland}

\author[0000-0002-1662-5756]{Piotr Wielg\'{o}rski}
\affiliation{Nicolaus Copernicus Astronomical Center, Polish Academy of Sciences, Bartycka 18, 00-716, Warsaw, Poland}

\author[0000-0003-1515-6107]{Bart{\l}omiej Zgirski}
\affiliation{Nicolaus Copernicus Astronomical Center, Polish Academy of Sciences, Bartycka 18, 00-716, Warsaw, Poland}

\author{M\'{o}nica Taormina}
\affiliation{Nicolaus Copernicus Astronomical Center, Polish Academy of Sciences, Bartycka 18, 00-716, Warsaw, Poland}

\author{Mradumay Sadh}
\affiliation{Nicolaus Copernicus Astronomical Center, Polish Academy of Sciences, Bartycka 18, 00-716, Warsaw, Poland}

\begin{abstract}
Deep near-infrared $J$- and $K$-band photometry of three Local Group dwarf spheroidal galaxies: Fornax, Carina, and Sculptor, is made available for the community. Until now, these data have only been used by the Araucaria Project to determine distances using the tip of the red giant and RR Lyrae stars. Now, we present the entire data collection in a form of a database, consisting of accurate $J$- and $K$-band magnitudes, sky coordinates, ellipticity measurements, and timestamps of observations, complemented by stars' loci in their reference images. Depth of our photometry reaches about 22\,mag at 5$\sigma$ level, and is comparable to NIR surveys, like the UKIRT Infrared Deep Sky Survey (UKIDSS) or the VISTA Hemisphere Survey (VHS). Small overlap with VHS and no overlap with UKIDSS makes our database a unique source of quality photometry.
\end{abstract}


\section{Introduction}
\label{sec:intro}

During the course of the Araucaria Project \citep{gieren05} we have collected high quality near-infrared (NIR) $J$- and $K$-band photometry for eleven galaxies in the Local Group, and four galaxies in the Sculptor Group, in order to calibrate the cosmic distance scale with primary distance indicators, i.a. classical Cepheids \citep{pietrzynski06,gieren13,zgirski17}, RR Lyrae stars \citep[RRL;][]{szewczyk08,szewczyk09,karczmarek15,karczmarek17}, the tip of the red giant branch stars \citep[TRGB;][]{pietrzynski09,gorski11}, late-type eclipsing binaries \citep{graczyk14,pietrzynski19}. The bulk of these data was collected with world-class ESO facilities over the course of 20 years, but only small part was selected to determine distances, leaving the vast rest of it unpublished and virtually idle.

Although these data are publicly available in their raw form in the ESO archive, the toil of downloading, reducing, performing profile photometry, and calibrating magnitudes to standard photometric system may discourage their use. In order to meet ever growing need for public and easy accessible data, we publish the final products in a form of a database available online.

We decided to publish our entire data collection in parts, due to several reasons. Our data were collected in either single- or multi-epoch mode of observations, depending on the method utilized to determine distances. For example, classical Cepheids require multiple observations to cover their light curves, for the calculation of accurate mean magnitudes, while observations of TRGB stars can be accomplished with just one exposure per field. In response to these modes of observations, we split our data collection into single- and multi-epoch data sets. Next, we made a further division regarding distances, i.e. we separated nearby galaxies in the Local Group, like the Magellanic Clouds, Carina, Fornax and Sculptor, from the rest of more distant ones (IC\,1613, M33, NGC\,3109, NGC\,6822, WLM), and from galaxies of the Sculptor Group (NGC\,55, NGC\,247, NGC\,300, NGC\,7793). 
As a result, our collection has four parts: I. Carina, Fornax and Sculptor galaxies, for their single-epoch photometry and short distances; II. The Magellanic Clouds, for their multi-epoch photometry and short distances; III. Distant galaxies in the Local Group; IV. Galaxies in the Sculptor Group. This paper includes part I, and presents the data collection for the dwarf spheroidal galaxies Carina, Fornax and Sculptor.

While these galaxies are devoid of Classical Cepheids, they do host populations of other standard candles, like RRL and TRGB stars, which have proved useful in precise and reliable distance determinations \citep{pietrzynski03,pietrzynski08,pietrzynski09,gorski11,karczmarek15,karczmarek17}. All three galaxies were scanned by the Two Micron All-Sky Survey \citep[2MASS;][]{TWOMASSdb03,TWOMASS06}, and only two of them -- Carina and Sculptor -- by the Deep Near-Infrared Survey of the Southern Sky \citep[DENIS;][]{DENIS98,DENISdb05}. Deep and high quality photometry was obtained by the VISTA Hemisphere Survey \citep[VHS;][]{VHS13,VHSdb19} for Carina, but not for Fornax nor Sculptor. While the detailed comparison of these surveys and our database can be found in Sec. \ref{sec:comparisonsurveys}, it is apparent that homogeneous and cohesive deep NIR photometry for all three galaxies has not been available in the literature. 
The database we publish is a product of reduction, photometry and calibration routines, which have been consistently utilized all throughout the Araucaria Project, guaranteeing quality and homogeneity of our data.

Unlike Fornax and Sculptor, Carina is located at a low galactic latitude of $b\approx-22^\circ$, relatively close to the Milky Way's disc. As a result, its color-magnitude diagram (CMD) is heavily contaminated by Milky Way stars, which in turn impedes the statistical analysis of Carina's stellar populations. In this paper we introduce a simple, auxiliary algorithm that finds and marks non-members of the Carina, Sculptor and Fornax galaxies, decreasing the foreground contamination in their CMDs.

All data that were collected for the Carina, Fornax and Sculptor galaxies during the course of the Araucaria Project, are described in section \ref{sec:observations} and outlined in appendix \ref{app:nightlog}. Data reduction, photometry and calibration are detailed in section \ref{sec:redphotcal}. Section \ref{sec:refinement} covers a number of additional steps we took to refine the data in order to boost their usefulness. In section \ref{sec:results} we describe our database in detail, compare it to other NIR surveys, and inform how to access it. Section \ref{sec:conclusions} concludes the paper with a recall of the main characteristics of our database.

\section{Observations}
\label{sec:observations}

\begin{figure}[ht]
    \includegraphics[width=0.84\hsize]{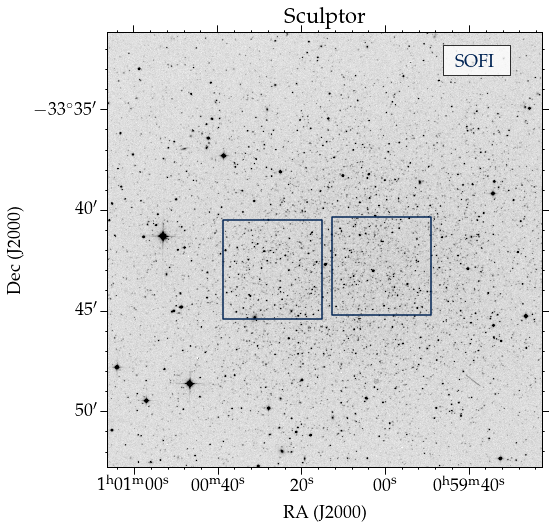}\vspace{2mm}\\
    \includegraphics[width=0.84\hsize]{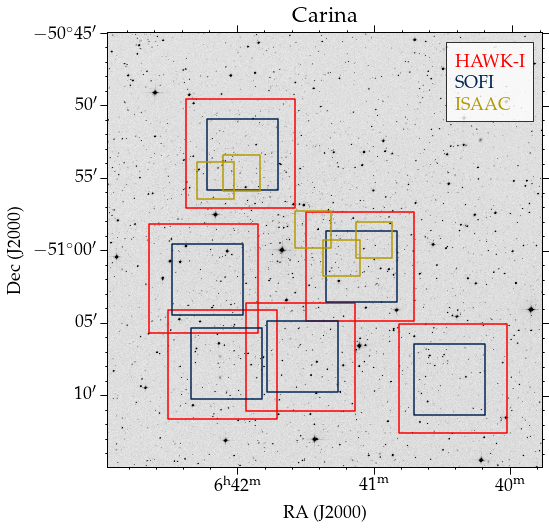}\vspace{2mm}\\
    \includegraphics[width=0.84\hsize]{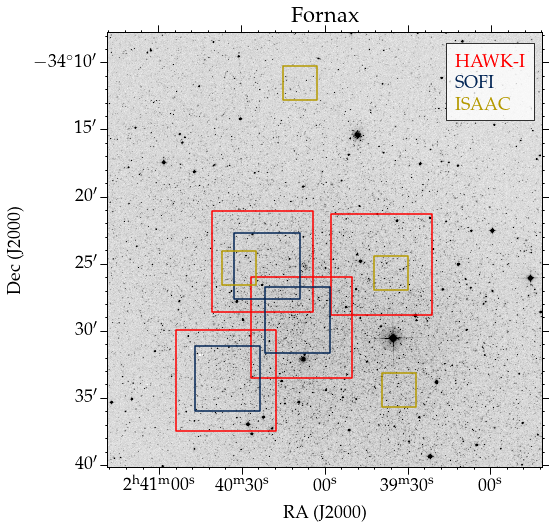}
    \caption{Sculptor, Carina, and Fornax galaxies with instrument footprints, color coded by the instrument, marked on the DSS-2 infrared plates.}
    \label{fig:FoVs}
\end{figure}

The NIR data were collected between years 2001 and 2013 as a part of the Araucaria Project, with three NIR cameras: SOFI \citep{sofi_moorwood98} at the New Technology Telescope (NTT) in La Silla Observatory, ISAAC \citep{isaac_moorwood98} and \hawki\ \citep{hawki_pirard04,hawki_casali06,hawki_kissler-patig08} at the Very Large Telescope (VLT) in Paranal Observatory. These instruments have high spatial resolutions and fields of view: $4.9' \times 4.9'$ and $0.288''\,\mathrm{pix}^{-1}$ (SOFI), $7.5' \times 7.5'$ and $0.106''\,\mathrm{pix}^{-1}$ (\hawki), $2.5' \times 2.5'$ and $0.148''\,\mathrm{pix}^{-1}$ (ISAAC). The locations of target fields in Fornax, Carina and Sculptor galaxies are presented in Figure \ref{fig:FoVs}. The fields were observed in $J$ and $K_\mathrm{s}$ (from hereon denoted as $K$) passbands, under photometric conditions. The observations were intended to be single-epoch, although a few stars have multiple data points, as they come from overlapping fields or fields observed more than once.

In order to account for frequent sky-level variations, which are especially strong in the NIR domain, the observations were carried out using jittering, i.e. consecutive exposures of a given field were randomly shifted by $20''$ with respect to the initial position. The number of consecutive jitter exposures varied with filters, instruments, and targets; generally, for SOFI we took between 19 and 21 exposures in $J$-band and 25-55 exposures in $K$-band, for \hawki\ we took between 15 and 18 exposures in $J$-band and 26-45 in $K$-band, and for ISAAC we took between 4 and 20 exposures in $J$-band and 24-26 in $K$-band. All jitter exposures were shorter or equal to one minute, and in case of 1-minute exposures, the saturation in $K$-band was avoided by aggregating the light from five 12-second sub-integrations (DITs) into one 1-minute exposure. In $J$-band 1-minute exposures were obtained from two 30-second sub-integrations. Total integration time per field, as a product of integration time and number of jitter exposures, is summarized in an observational journal (Appendix \ref{app:nightlog}, Table \ref{tab:nightlog}), together with coordinates of centers of fields, dates of observations, seeing conditions, filters and instruments' names.

Together with science targets, we observed 10-20 standard stars from the United Kingdom Infrared Telescope list \citep[UKIRT;][]{hawarden01} at different air masses and spanning a broad range in color, in order to perform absolute calibration of our science targets.

\section{Data reduction, photometry, and calibration}
\label{sec:redphotcal}
Reduction routines for ISAAC and SOFI were identical, and differed slightly from the reduction routine for \hawki. Although the routines were already detailed by \citet{pietrzynski02} and \citet{karczmarek15}, here we recapitulate them step by step, showing their similarities and differences.

Basic reductions (dark correction, flat fielding, cosmic rays removal, and bad pixel correction) of data from SOFI and ISAAC were performed on FITS files using the IRAF package \citep{iraf_tody86}. In the NIR domain the sky brightness is significant, and therefore background removal is crucial to calculate the brightness of celestial objects. The sky was subtracted from the images in a two-step process, using the xdimsum IRAF package. The first background subtraction was performed using the median of the adjacent science images. Then, all stars were masked, so that their contribution was ignored, and the background was calculated and subtracted for the second time. The individual images were then corrected for the shifts caused by jittering and stacked into a final image.

Reductions of \hawki\ data were performed entirely using a module of ESO Recipe Execution Tool (EsoRex), designed to handle MEF (Multiple Extention FITS) files produced by the \hawki\ instrument. Basic reductions and two-step background subtraction were executed in an identical manner as described above. The next procedure, unique to the \hawki\ reduction routine, minimized the effect of any distortions caused by atmospheric refraction or non-planar surface of detectors. Lastly, frames were corrected for the shifts caused by jittering, and stacked into a final image, exactly as in the reduction routines for SOFI and ISAAC.

In case of standard stars, which were captured on single frames (no jittering), the procedure leading to the final product consisted of fewer steps: basic calibration (dark correction, flat fielding), single-step sky subtraction, and distortion correction in case of \hawki\ images.

\hawki\ images, so far in a form of MEF files, were next split into four FITS files, which unified our set of data and allowed us to perform profile photometry in a homogeneous manner, described below.

Point spread function (PSF) photometry was carried out with standalone DAOPHOT and ALLSTAR programs \citep{stetson87}. About 10-20 relatively bright and isolated stars were selected visually, and the first PSF model was derived from them. Then, the PSF model was improved iteratively by subtracting all neighbouring stars and re-calculating the PSF model. Generally, after three such iterations no further improvement was noted, and the corresponding PSF model was adopted. In order to prepare our data to be transformed to standard photometric system, aperture corrections were derived for each frame. This was done by calculating aperture photometry of previously identified candidates for PSF calculations. The median from the aperture corrections derived for these stars was adopted as an aperture correction for a given frame. 

In the last step, the instrumental magnitudes were calibrated onto the UKIRT photometric system \citep{hawarden01}. To prepare our data for this calibration, aperture photometry on the standard stars was performed by choosing an aperture radius of 16 pixels for ISAAC, 12 pixels for SOFI, and 14 pixels for \hawki. Then, three coefficients of transformation: zero point, $z$, extinction, $k$, and color term, $c$, were calculated simultaneously in order to secure the best fit between instrumental aperture magnitudes of standard stars scaled to 1-second exposure times (lower case) and UKIRT magnitudes (upper case) of these standard stars, in $J$-, $K$-bands and $(J-K)$ color:
\begin{eqnarray}
    J-K &=& c_{JK} (j-k) + k_{JK} \chi + z_{JK} \nonumber \\
    J &=& j + c_{J} (J-K) + k_J \chi + z_J \nonumber \\
    K &=& k + c_{K} (J-K) + k_K \chi + z_K \nonumber
\end{eqnarray}
Such a transformation resulted in a unique set of nine coefficients ($c_{JK}$, $k_{JK}$, $z_{JK}$, $c_{J}$, $k_J$, $z_J$, $c_{K}$, $k_K$, $z_K$) for every night. 
Finally, instrumental magnitudes, transformation coefficients and aperture corrections, $r$, were used to calibrate instrumental magnitudes onto the UKIRT photometric system, following the formulas:
\begin{eqnarray}
    \label{eq:J}
    J &=& j + k_{J} \chi + c_{J} (j-k)  + z_J \nonumber \\
    &+& 2.5\log(t_{\mathrm{exp},J}) + r_J \\
    \label{eq:K}
    K &=& k + k_{K} \chi + c_{K} (j-k)  + z_K \nonumber \\
    &+& 2.5\log(t_{\mathrm{exp},K}) + r_K \\
    \label{eq:JK}
    (J-K) &=& (j-k) + k_{JK} \chi + c_{JK} (j-k)  + z_{JK} \nonumber \\
    &+& 2.5\log(t_{\mathrm{exp},J}/t_{\mathrm{exp},K}) + r_J - r_K
\end{eqnarray}
Although the $J$ magnitude can be derived from color $(J-K)$ and $K$ magnitude, an independent calculation of $J$ magnitude constitutes an additional check of the correctness of the transformation. The accuracy of the zero point of our photometry is estimated to be 0.02 mag.

In order to make an external check of our photometric zero point, the magnitudes of the bright stars in our fields were transformed
onto the Two Micron All Sky Survey (2MASS) photometric system, using equations from \citet{carpenter01}, and then compared with the 2MASS photometry \citep[see][]{pietrzynski08,karczmarek15,karczmarek17}. Additionally, for the Carina galaxy, we transformed our data to the photometric system of the Visible and Infrared Survey Telescope for Astronomy \citep[VISTA;][]{gonzalez-fernandez18} and performed an external check with the VISTA Hemisphere Survey (VHS); this check was not possible for Sculptor nor Fornax due to the lack of VHS coverage for these galaxies. All photometry checks are assembled in Table \ref{tab:zp_check}.

\begin{deluxetable}{lcccc}
\tablecaption{Checks of photometric zero points with near-infrared surveys.}
\label{tab:zp_check}
\tablehead{
\colhead{Galaxy} & \colhead{Survey} & 
\colhead{Zero point [mag]} & \colhead{Filter} & \colhead{Ref.}
} 
\startdata
Sculptor & 2MASS & $-0.03\pm0.13$ & $J$ & 1\\
Sculptor & 2MASS & $-0.01\pm0.11$ & $K$ & 1\\
Carina & 2MASS & ~~$0.02\pm0.05$ & $J$ & 2\\
Carina & 2MASS & ~~$0.04\pm0.07$ & $K$ & 2\\
Carina & VHS & $-0.02\pm0.05$ & $J$ & this work\\
Carina & VHS & $-0.03\pm0.06$ & $K$ & this work\\
Fornax & 2MASS & $-0.01\pm0.12$ & $J$ & 3\\
Fornax & 2MASS & ~~$0.01\pm0.11$ & $K$ & 3\\
\enddata
\tablerefs{(1) \citealt{pietrzynski08}; (2) \citealt{karczmarek15}; (3) \citealt{karczmarek17}.}
\end{deluxetable}

\begin{figure}[ht]
\centering
\includegraphics[width=\hsize]{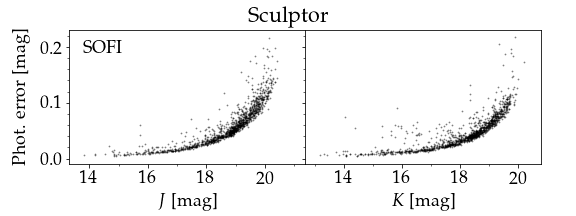}\\
\includegraphics[width=\hsize]{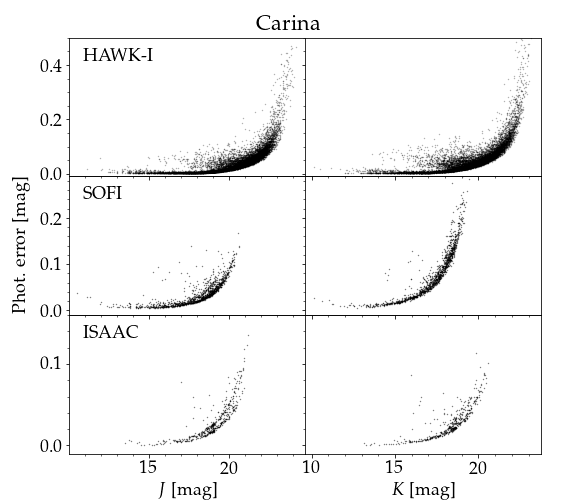}\\
\includegraphics[width=\hsize]{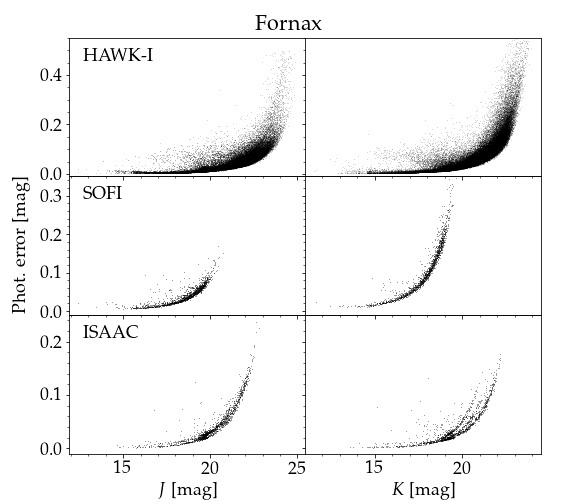}
\caption{Quality of photometry for Sculptor, Carina and Fornax galaxies, shown on magnitude error versus magnitude plot.}
\label{fig:magerrmag}
\end{figure}

The quality of our photometry can be evaluated in Figure \ref{fig:magerrmag} for all three galaxies. In case of Carina and Fornax, where data come from more instruments (SOFI, ISAAC and \hawki), separate images for every instrument present trends of growing photometric errors.

Lower photometry limits at levels of $5\sigma$ and $10\sigma$ are equivalent of signal-to-noise ratio (SNR) of 5 and 10, respectively. For Sculptor, they are 20.1; 19.7; 19.6; 19.2\,mag for $J_{5\sigma}$, $J_{10\sigma}$, $K_{5\sigma}$, $K_{10\sigma}$, respectively. For Carina, the same limits are 22.9; 22.2; 21.6; 21.2\,mag, while for Fornax they are 23.3; 22.5; 22.4; 21.7\,mag. Difference in liming magnitudes for these galaxies arise from different exposure times. A complementary way to assess the quality of photometry is to perform the completeness analysis with artificial star experiments -- this approach is detailed in Section \ref{sec:completeness}. 

Upper limits of photometry reflect the saturation limits for each detector. ISAAC and SOFI are linear as long as the counts level of the source and the sky is below 10\,000 \,ADU, 
for \hawki\ this limit is 30\,000\,ADU. The limits translate to about 14\,mag ($K$) and 15\,mag ($J$) for SOFI, 14.5\,mag ($K$) and 15.5\,mag ($J$) for ISAAC, and 15\,mag ($K$) and 16\,mag ($J$) for \hawki, meaning that a handful of the brightest stars in our data set might suffer from slightly inaccurate magnitude determinations caused by non-linearity.

Due to the jittering mode of observations, the edges of final images suffer from much lower signal-to-noise ratio. We decided to either trim the final images or to ignore this edge area, so that only stars located farther from the edge are included in the database. The edge area extends for 40 pixels from the image edge inward for SOFI and ISAAC and 80 pixels for \hawki.

\section{Quality check and completeness analysis}
\label{sec:refinement}
In order to provide the scientific community with the best experience of using our database, we included a number of parameters that elevate the usability of our data. 

\subsection{Ellipticity}

The information about ellipticity helps to evaluate whether an object is a point or extended source. In case of stars, which are expected to be round point sources and thus have zero ellipticity, ellipticity combined with magnitude error is a useful diagnostic of the quality of photometry.
For every object in our database ellipticities were calculated with SExtractor \citep{sextractor96}, separately for $J$- and $K$-band. In rare cases where ellipticity was not derived, its value was substituted with not a number, i.e. 'NaN'.

\subsection{Refinement of coordinates}
Sub-arcsec accuracy of the right ascension and declination for every target in our database was achieved using sky coordinates from the Gaia EDR3 catalog \citep{gaia18,gaia20}. This step was necessary because the coordinates of the center of every FITS file were accurate to only about 10\,arcsec, which impedes the proper localization of targets, especially in dense fields.

\begin{figure}
    \centering
    \includegraphics[width=\hsize]{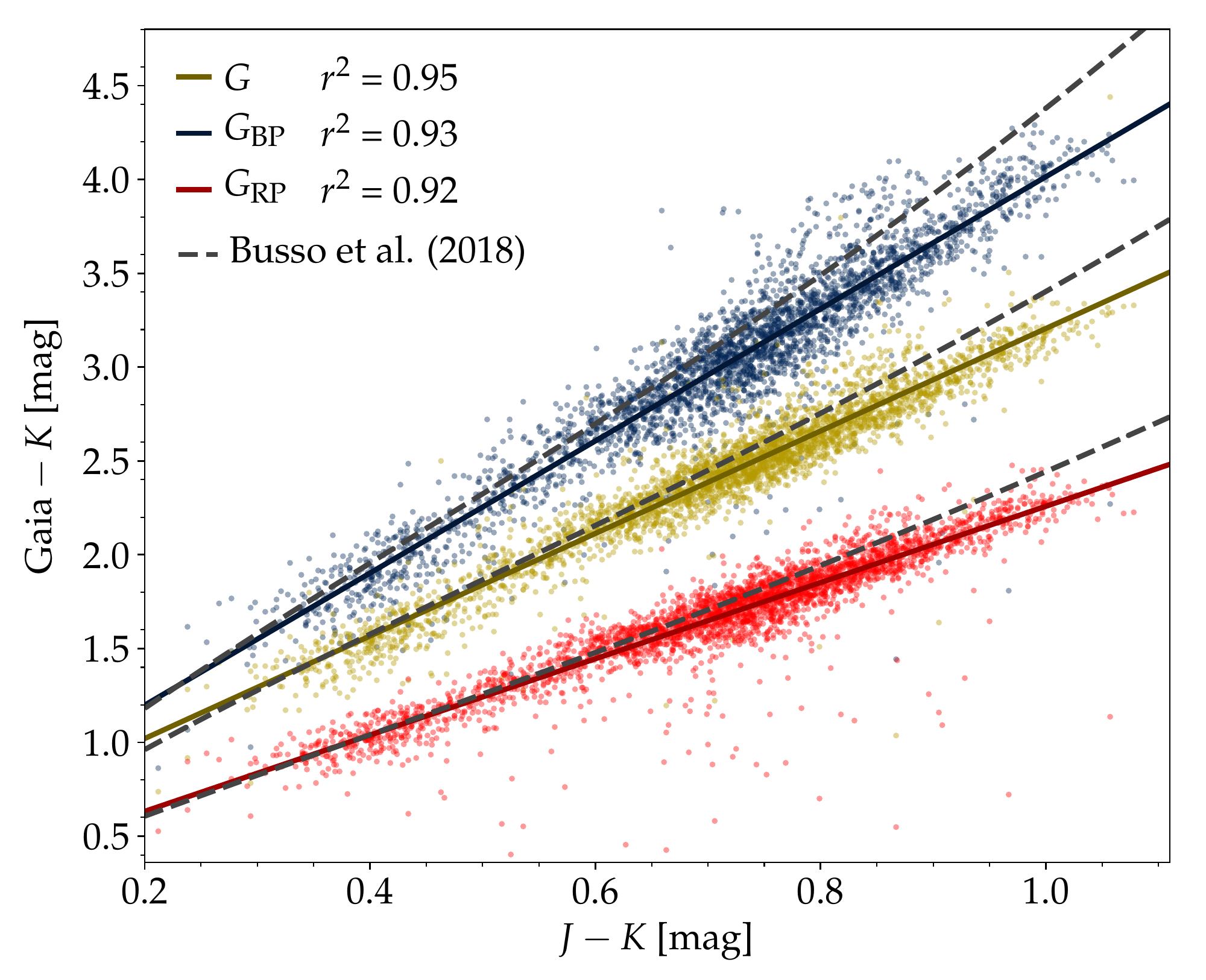}
    \caption{Photometric relationships obtained using Gaia EDR3 data with $G < 19.8$\,mag, $G_\mathrm{BP} < 20.2$\,mag, $G_\mathrm{RP} < 19.0$\,mag, $(G_\mathrm{BP}-G_\mathrm{RP}) \in [0.0, 1.9 ]$\,mag, cross-matched with our data in the UKIRT photometric system. Gaia passbands are color coded, and the coefficient of determination, $r^2$, represents the goodness of linear fit. For comparison, grey dashed lines show transformations of \citet{busso18} for Gaia DR2 magnitudes $G<13$\,mag cross-matched with 2MASS catalog.}
    \label{fig:photsyst}
\end{figure}

We used the coordinates of the center of every FITS image and queried the Gaia EDR3 archive for all stars inside a box of the same size as the field of view of our image. We trimmed the set of Gaia stars allowing only for objects with brightnesses within the magnitude range of stars in our image. Because our targets have $J$- and $K$ magnitudes, and Gaia's targets have $G$, $G_\mathrm{BP}$, $G_\mathrm{RP}$ magnitudes, a transformation between these photometric systems was necessary. For this transformation we used fields in Carina and Fornax, because of their moderate density, that allowed an easy cross-match with Gaia EDR3 catalog. Relations between Gaia and UKIRT magnitudes are presented in Figure \ref{fig:photsyst} and with formulas \ref{eq:gaiaukirtG}, \ref{eq:gaiaukirtB}, \ref{eq:gaiaukirtR}:
\begin{eqnarray}
    \label{eq:gaiaukirtG}
    G&=&K+ 0.474 (\pm 0.012) \nonumber\\
    &&+2.732 (\pm 0.016) (J-K) \\
    \label{eq:gaiaukirtB}
    G_\mathrm{BP}&=&K+ 0.494(\pm0.017) \nonumber\\
    &&+3.521(\pm0.023)(J-K) \\
    \label{eq:gaiaukirtR}
    G_\mathrm{RP}&=&K+ 0.226(\pm0.011) \nonumber\\
    &&+2.033(\pm0.015)(J-K)
\end{eqnarray}

Our transformations between the Gaia and UKIRT photometric systems are suitable for Gaia stars with magnitudes $G < 19.8$\,mag, $G_\mathrm{BP} < 20.2$\,mag, $G_\mathrm{RP} < 19.0$\,mag, and color $(G_\mathrm{BP}-G_\mathrm{RP}) \in [0.0, 1.9 ]$\,mag.
Similar work was done by \citet{busso18} for Gaia magnitudes $G<13$\,mag cross-matched with 2MASS catalog. Their result agrees with our fit quite well, even though \citet{busso18} fitted a $4^\mathrm{th}$ degree polynomial, and we fitted a line.

\begin{figure}
    \centering
    \includegraphics[width=\hsize]{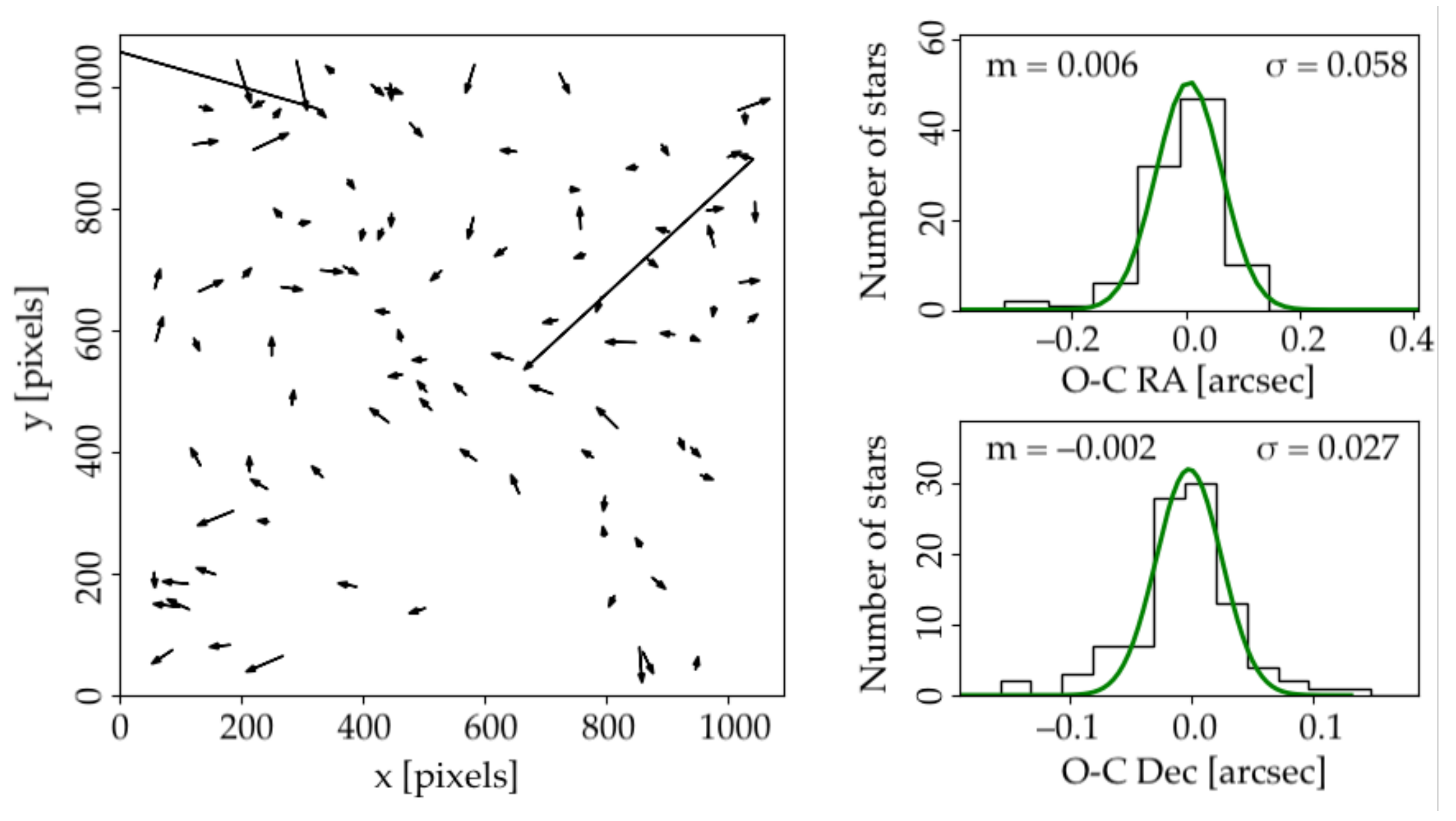}
    \caption{Exemplary cross-match between original locations of Gaia stars and locations of our stars, transformed to Gaia coordinates, for Car-F04\_131127\_K field. Vectors point from original Gaia to transformed coordinates and are magnified 100 times for clarity. Distributions in RA and Dec show the mean and the standard deviation from the mean difference between Gaia (O) and transformed (C) coordinates.}
    \label{fig:coords_refinement}
\end{figure}

The next step towards the sub-arcsec accuracy was to cross-match our targets from a given image with targets from Gaia subset, that now contained only stars of relevant magnitudes. This step was done using ASTROALIGN \citep{astroalign20}, a python module that aligns two astronomical images (or data sets) by finding similar 3-point asterisms and estimating the affine transformation between them. This tool is especially useful for images with incomplete or inaccurate WCS information. The process is completely automatic for fields of moderate density, and for more dense or scarce fields it requires additional help in a form of pre-defined (e.g. eye-picked) pairs of stars, that are used to perform the transformation. The outstanding accuracy of the transformation is shown in Figure \ref{fig:coords_refinement} for an exemplary field of Car04\_131127\_1\_K. Vectors on a pixel plane present a qualitative result; they point from original Gaia to transformed coordinates and are magnified 100-fold for clarity. Distributions in right ascension and declination give the quantitative information of the mean and the standard deviation from the mean difference between Gaia and transformed coordinates. Even though a few large vectors indicate improper match, the overall transformation is excellent. 

After the transformation, new and accurate WCS information was saved in FITS files. A visual inspection of stars' new locations was done by comparing our FITS files with images from the Digitized Sky Survey (DSS) in ALADIN sky atlas \citep{aladin00}.

\subsection{Galaxy membership}

\begin{figure}[t]
    \centering
    \vspace{3mm}
    \includegraphics[width=\hsize]{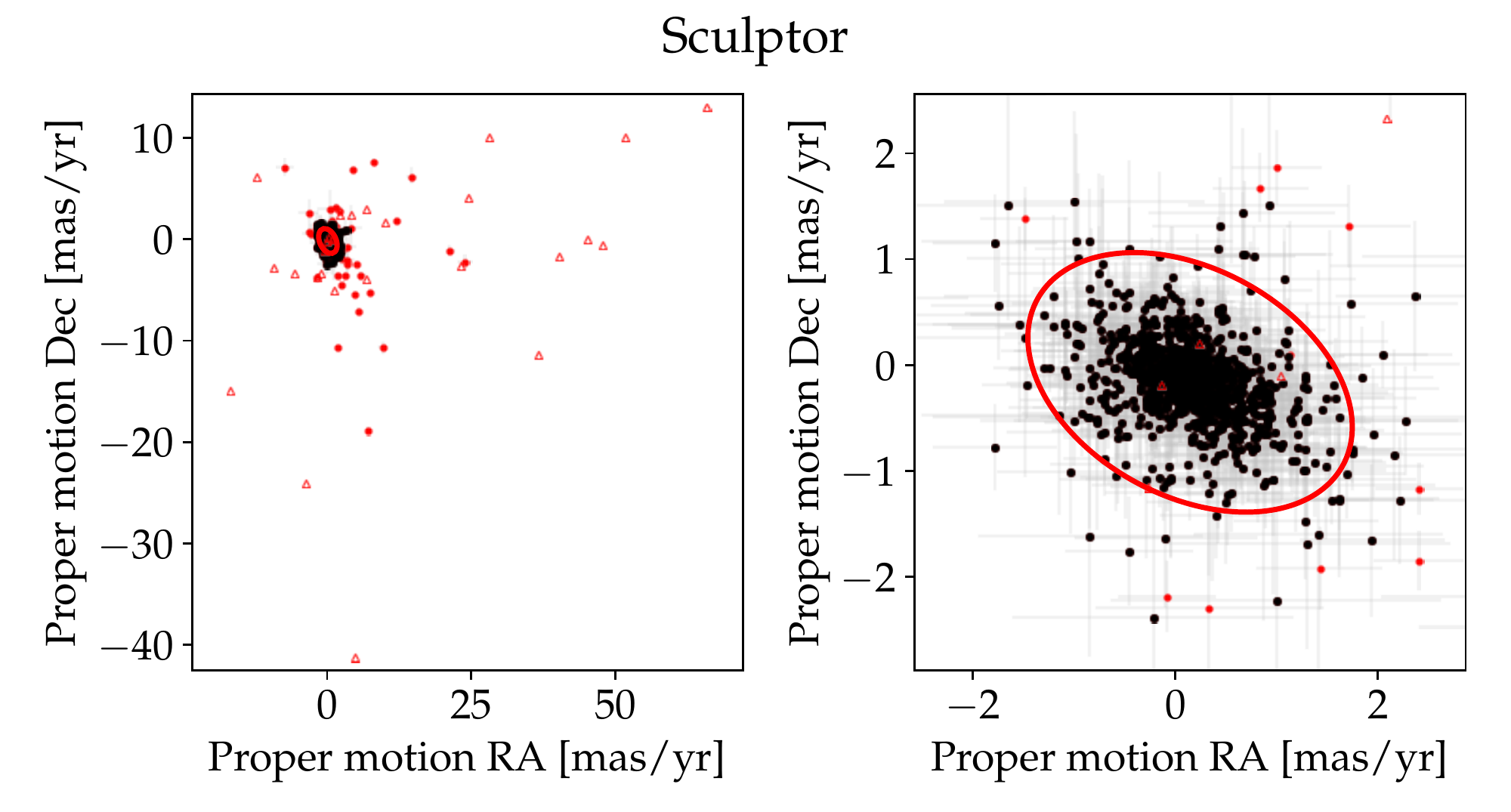}
    \includegraphics[width=\hsize]{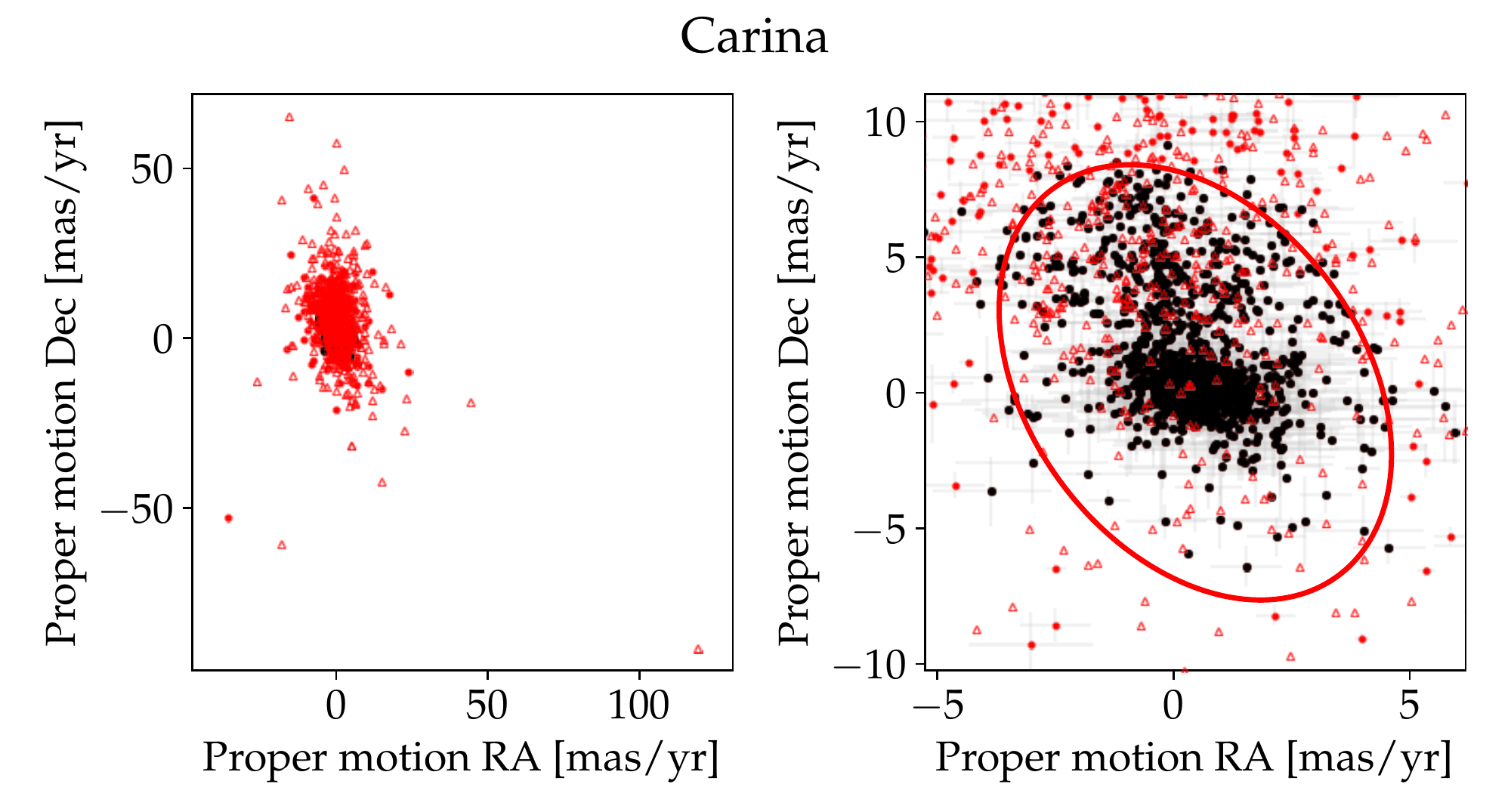}
    \includegraphics[width=\hsize]{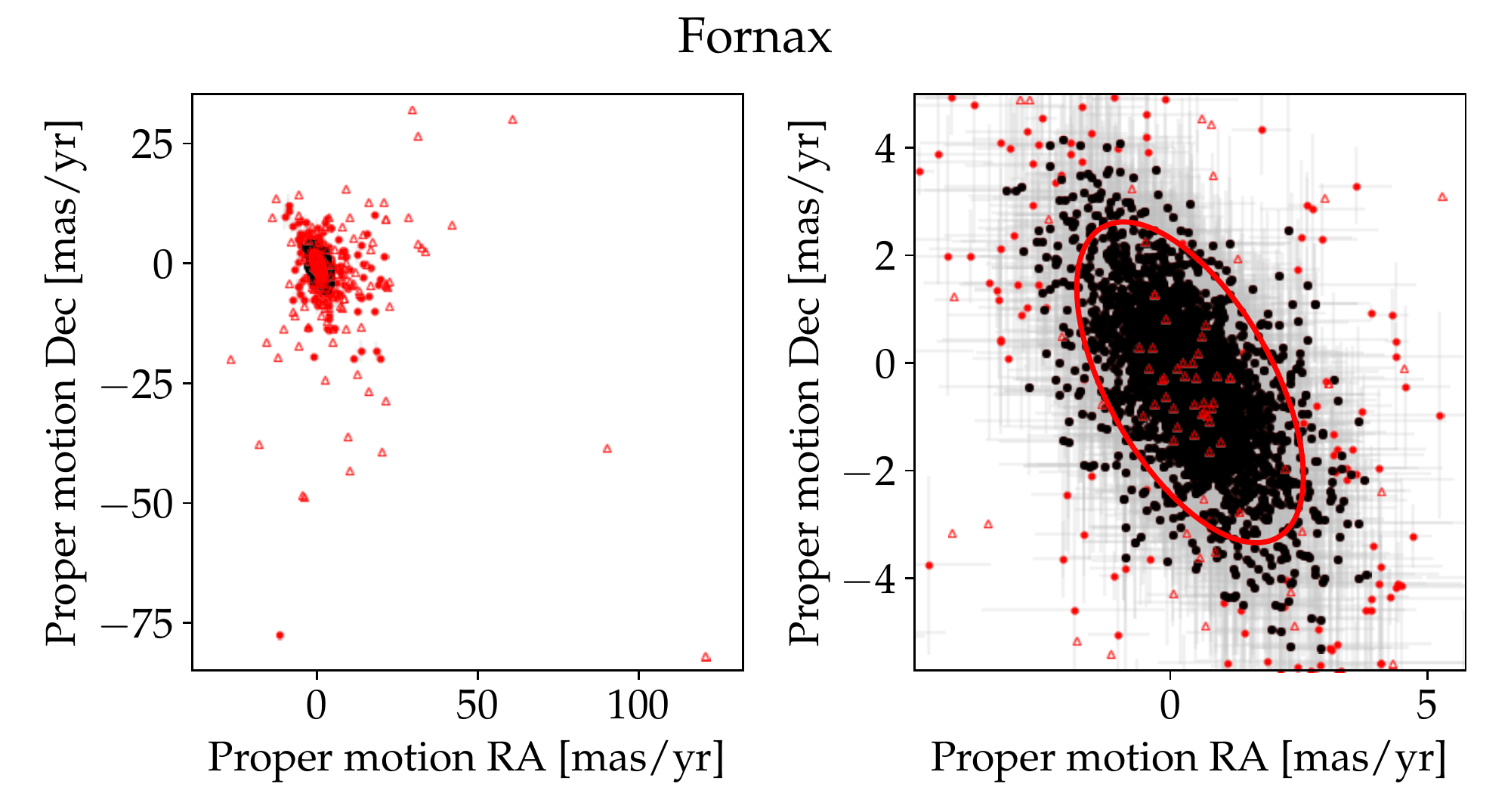}
    \caption{Vector point diagram of Gaia EDR3 stars from Sculptor, Carina, and Fornax (left: entire sample, right: zoomed). Red triangles mark foreground stars based on their high parallax values. Red circles mark foreground stars based on their proper motions. Black circles mark non-foreground stars, which are contained (within their uncertainties) in 95\% confidence ellipse.}\label{fig:parallax_propmot}
\end{figure}

We define the 'frg' parameter that informs if a star is a foreground object that does not belong to a target galaxy. This parameter is especially useful in case of Carina, which is substantially contaminated by foreground stars. Such stars are expected to have large values of parallax and proper motions, non-coinciding with the mean proper motion of the galaxy. 
We utilized the information about proper motions and parallaxes from Gaia online archive for common stars in our fields and Gaia EDR3 catalog. 
We assigned one of three values to the parameter 'frg': 1 if the object is a foreground star, 0 if the object is not a foreground star, or 'Nan' if its membership could not be established. 'Nan' values were set in two cases: i) faint stars, which were below Gaia detection threshold, and therefore could not be cross-matched; ii) stars with negative parallaxes. For the rest of objects, we compared their parallaxes with expected parallaxes of stars belonging to Sculptor, Carina, and Fornax galaxies, based on well-defined and precise distances from NIR period-luminosity-metallicity of RR Lyrae stars, reported by \citet{pietrzynski08} and \citet{karczmarek15,karczmarek17}, and marked as non-members those with parallaxes significantly larger (beyond 3$\sigma$) than the parallaxes of target galaxies. Remaining stars were plotted on a proper-motion vector point diagram (VPD), shown in Figure \ref{fig:parallax_propmot} for Sculptor, Carina and Fornax. Points follow bivariate normal distribution, and the representation of this distribution, that encapsulates 95\% of all stars, is an elliptical area centered at the mean proper motion, averaged over all stars, whose parallaxes agree (within 3$\sigma$) with the galaxy parallax. Objects outside the ellipse are marked as foreground stars ($\mbox{fgr}=1$), while objects inside the ellipse (within the uncertainties) are marked as non-foreground ($\mbox{fgr}=0$). While this approach is not sufficient to identify galaxy members, it is highly reliable in detecting foreground stars. Its caveat is modest overlap between Gaia and our database; in the case of Sculptor, 62.5\% of our stars were found in Gaia catalog, for Carina this value was 23.5\%, and for Fornax -- only 16.7\%. This overlap is magnitude-dependent, meaning that for all our objects brighter than $J=14.5$\,mag the coverage with Gaia catalogue is complete, i.e. 100\%, then decreases linearly to 65\% for $J=19.7$\,mag, when suddenly drops to 20\% and remains constant for all fainter objects. Because of this limitation, only 3.8\% of stars in Sculptor, 8.5\% in Carina, and 1.2\% in Fornax, were classified as foreground ($\mbox{fgr}=1$). 

\subsection{Completeness}
\label{sec:completeness}

\begin{figure}[ht]
    \centering
    \includegraphics[width=0.81\hsize]{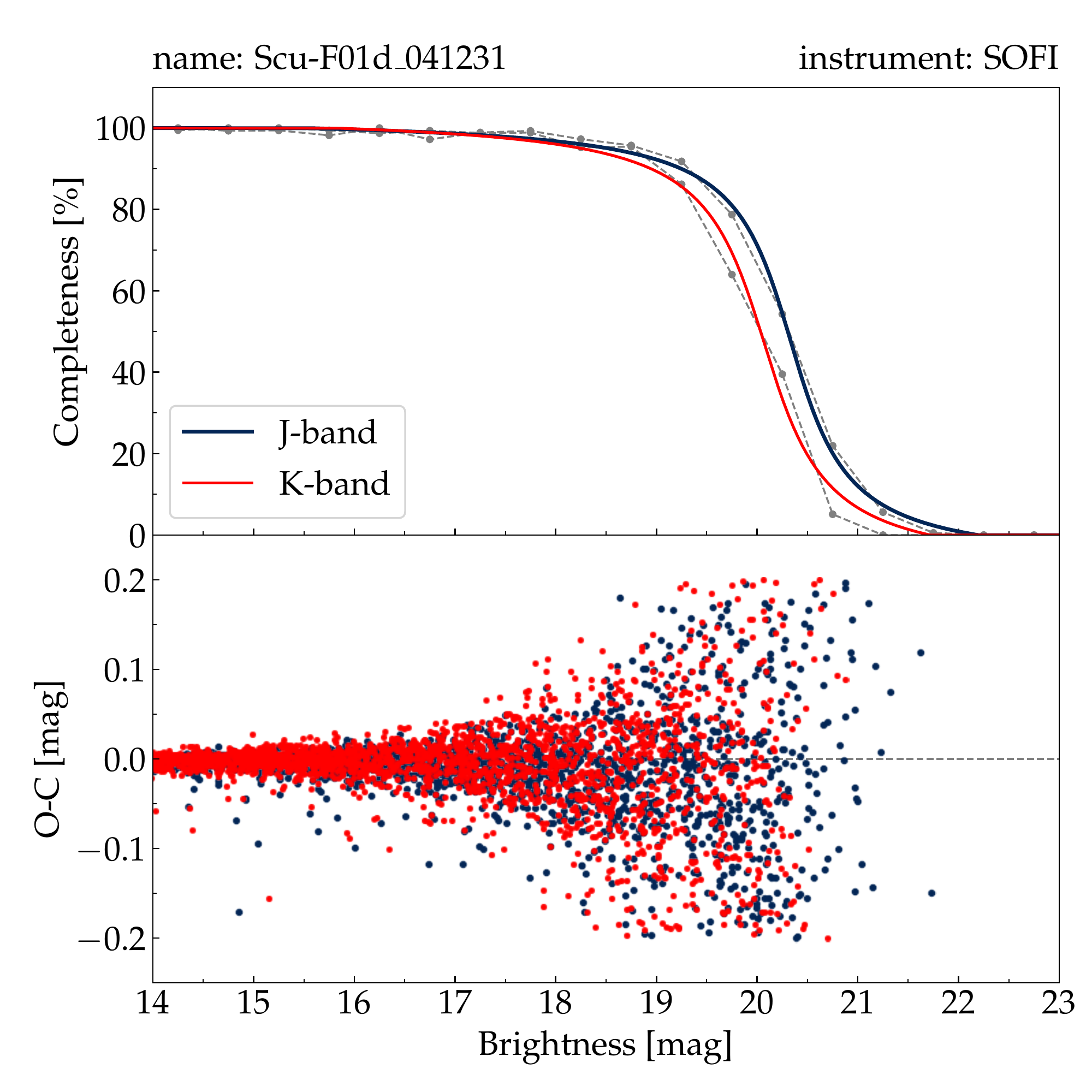}\\
    \includegraphics[width=0.81\hsize]{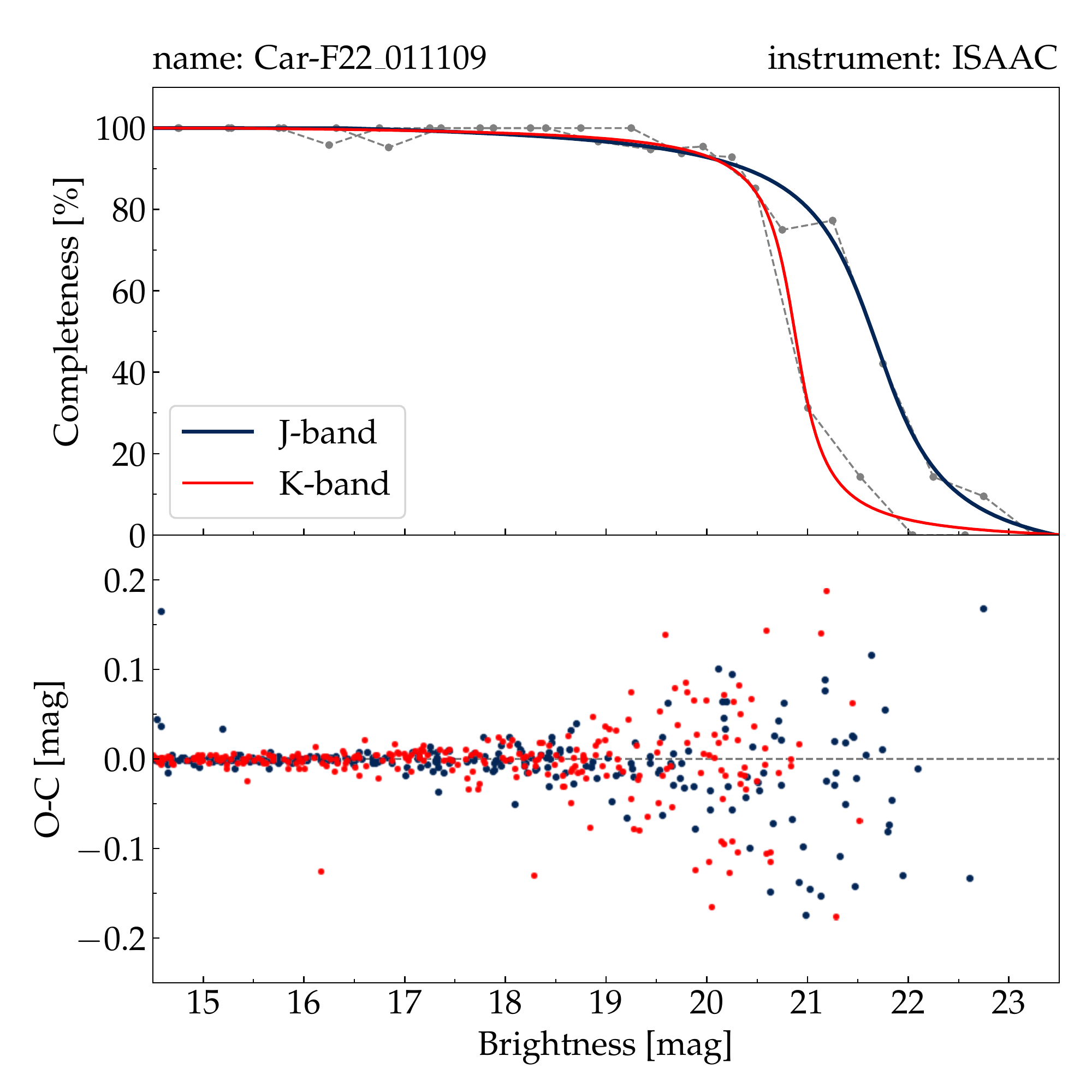}\\
    \includegraphics[width=0.81\hsize]{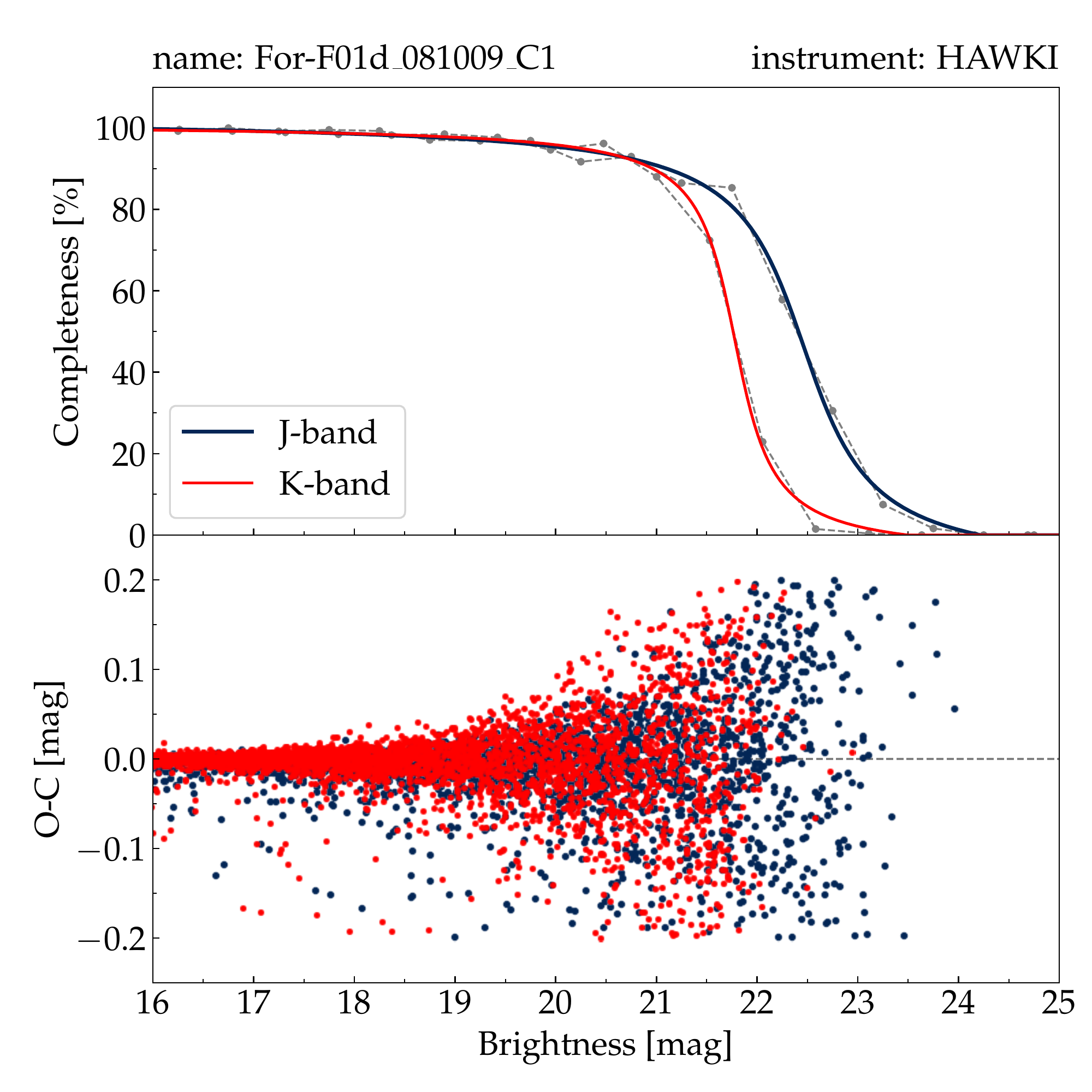}
    \caption{Completeness and photometric error analysis for exemplary fields in the Sculptor, Carina, and Fornax galaxies as a result of artificial-star experiments. Figures for all fields are available on Zenodo and Araucaria website.}
    \label{fig:completeness}
\end{figure}

Artificial-star experiments, in which new stars were added to the processed images using the DAOPHOT ADDSTAR routine, were run to estimate completeness and assess the scatter due to photometric errors. The artificial stars were assigned brightnesses in the range of instrumental magnitudes of stars detected in a given image. Each image had their unique PSF computed, as described in section \ref{sec:redphotcal}, and this PSF was used to create artificial stars in each image. 
The number of added artificial stars was calculated individually for every image, but was always limited to 5-10\% of all stars in the image, in order to avoid the blending and crowding effects. This exercise was done separately for $J$ and $K$ images, and the resulting instrumental magnitudes of artificial stars were then transformed to the UKIRT photometric system. Figure \ref{fig:completeness} presents exemplary completeness plots for one field in Carina, Fornax and Sculptor galaxies, while the entire collection of plots for every field is available online. Together with plots, we make available in a digitized form Table \ref{tab:compl_For}, where we present the limiting magnitudes for completeness levels of 99, 95, 90, 75 and 50\% for every field in $J$ and $K$ filter. 

\begin{deluxetable*}{lccccccc}
\tablecaption{Completeness estimation for Fornax fields. }\label{tab:compl_For}
\tablehead{
\colhead{Name of reference} & \colhead{Instrument} & \colhead{Filter} &
\colhead{Compl. 99\%} & \colhead{Compl. 95\%} & \colhead{Compl. 90\%} & 
\colhead{Compl. 75\%} & \colhead{Compl. 50\%}\\
\colhead{image\tablenotemark{a}} & \colhead{} & \colhead{} &
\colhead{[mag]} & \colhead{[mag]} & \colhead{[mag]} & 
\colhead{[mag]} & \colhead{[mag]}
} 
    \startdata
    For-F04\_131127             & SOFI   & J & 15.5 & 19.0 & 19.7 & 20.3 & 20.5 \\
    For-F04\_131127             & SOFI   & K & 15.5 & 17.3 & 17.8 & 18.3 & 18.5 \\
    For-F04d\_081009\_C1  & HAWK-I & J & 18.8 & 21.2 & 22.0 & 22.8 & 23.2 \\
    For-F04d\_081009\_C1  & HAWK-I & K & 18.2 & 20.7 & 21.4 & 21.9 & 22.2 \\
    For-F04d\_081009\_C2  & HAWK-I & J & 16.0 & 20.3 & 21.5 & 22.5 & 23.0 \\
    For-F04d\_081009\_C2  & HAWK-I & K & 16.0 & 19.8 & 20.9 & 21.6 & 22.0 \\
    For-F04d\_081009\_C3  & HAWK-I & J & 18.0 & 20.4 & 21.4 & 22.4 & 23.0 \\
    For-F04d\_081009\_C3  & HAWK-I & K & 16.0 & 20.4 & 21.3 & 21.9 & 22.2 \\
    For-F04d\_081009\_C4  & HAWK-I & J & 18.7 & 21.3 & 22.1 & 22.8 & 23.1 \\
    For-F04d\_081009\_C4  & HAWK-I & K & 18.7 & 20.9 & 21.4 & 21.8 & 22.1 \\
    For-F11\_011110             & ISAAC  & J & 14.5 & 18.1 & 19.3 & 20.3 & 20.8 \\
    For-F11\_011110             & ISAAC  & K & 14.5 & 18.0 & 19.3 & 20.0 & 20.3 \\
    \enddata
\tablecomments{A complete table is available on Zenodo \citep{karczmarek20db} and Araucaria website. A portion is shown here for guidance regarding its form and content. Analogous tables are available online for Carina and Sculptor dwarf galaxies.}
\tablenotetext{a}{Each name consists of a galaxy name followed by a field number with a suffix 'd' in case of deeper photometry, date of observations in format YYMMDD, and information about chip number in case of HAWK-I images.}
\end{deluxetable*}

Mind, that the database consists only of stars that have both $J$ and $K$ magnitudes, and $K$-band completeness magnitudes are systematically shifted towards brighter stars, relative to the $J$-band magnitudes. This means that some stars, that were not detected in $K$ images and have only $J$ magnitudes,  were not included in our database. 

\section{Results}
\label{sec:results}

For each of thee dwarf galaxies: Sculptor, Carina, Fornax, we make available for the community a database containing the following information:
\begin{itemize}
    \setlength\itemsep{-0.3em}
    \item[(1)] identification number for every star,
    \item[(2)] right ascension (in degree units),
    \item[(3)] declination (in degree units),
    \item[(4)] $J$-band magnitude in the UKIRT photometric system,
    \item[(5)] uncertainty of $J$-band magnitude calculated by DAOPHOT (in magnitude units),
    \item[(6)] $K$-band magnitude in the UKIRT photometric system,
    \item[(7)] uncertainty of $K$-band magnitude calculated by DAOPHOT (in magnitude units),
    \item[(8)] ellipticity in $J$-band calculated by SExtractor, 'NaN' if not found,
    \item[(9)] ellipticity in $K$-band calculated by SExtractor, 'NaN' if not found,
    \item[(10)] Julian Date of $J$-band observations,
    \item[(11)] Julian Date of $K$-band observations,
    \item[(12)] foreground star (1 means foreground, 0 means non-foreground, 'NaN' means inconclusive),
    \item[(13)] x coordinate of star's centroid in $J$ image (in pixel units),
    \item[(14)] y coordinate of star's centroid in $J$ image (in pixel units),
    \item[(15)] x coordinate of star's centroid in $K$ image (in pixel units),
    \item[(16)] y coordinate of star's centroid in $K$ image (in pixel units),
    \item[(17)] name of instrument that took the images,
    \item[(18)] name of reference $J$ image, to which x,\,y centroids refer,
    \item[(19)] name of reference $K$ image, to which x,\,y centroids refer.
\end{itemize}

Above information is presented in Table \ref{tab:db} for first ten stars in the Carina galaxy, while the entire database for Carina, as well as for Fornax and Sculptor, is available online. In total, we present 1\,252 entries for Sculptor, 14\,597 entries for Carina, and 83\,716 entries for Fornax. The measurements are mostly single-epoch, but some stars have two or more measurements if they were located in overlapping fields or if the field was observed more than once. Multiple measurements were gathered regarding their magnitudes and coordinates, and were assigned the same unique identification number.

Additionally to the database, we make available also the FITS files, that can serve as finding charts, but also as stand-alone scientific images. As such, they can be processed in order to obtain magnitudes of the quality presented in this paper.

\begin{figure}[th!]
    \centering
    \includegraphics[width=0.79\hsize]{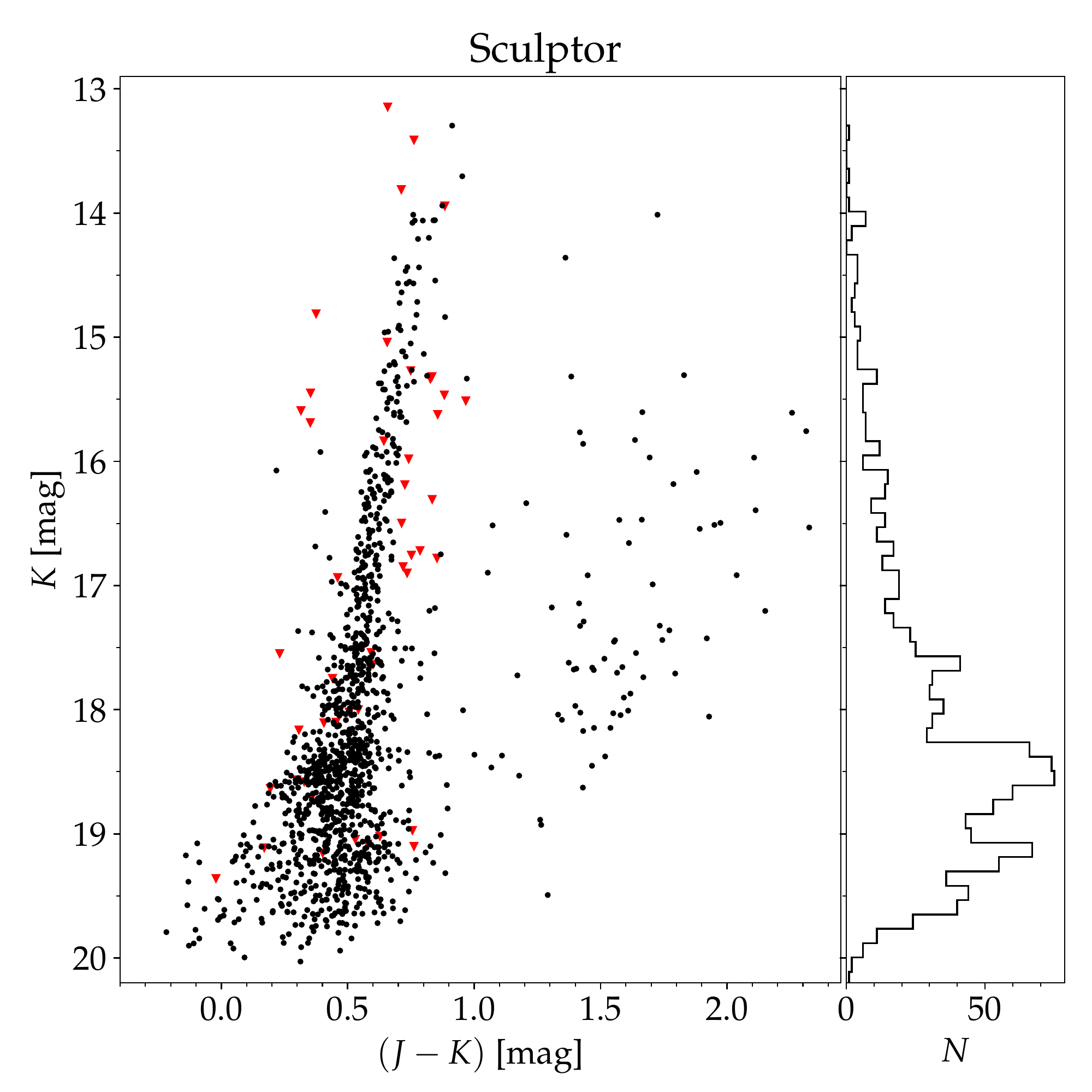}\\
    \includegraphics[width=0.79\hsize]{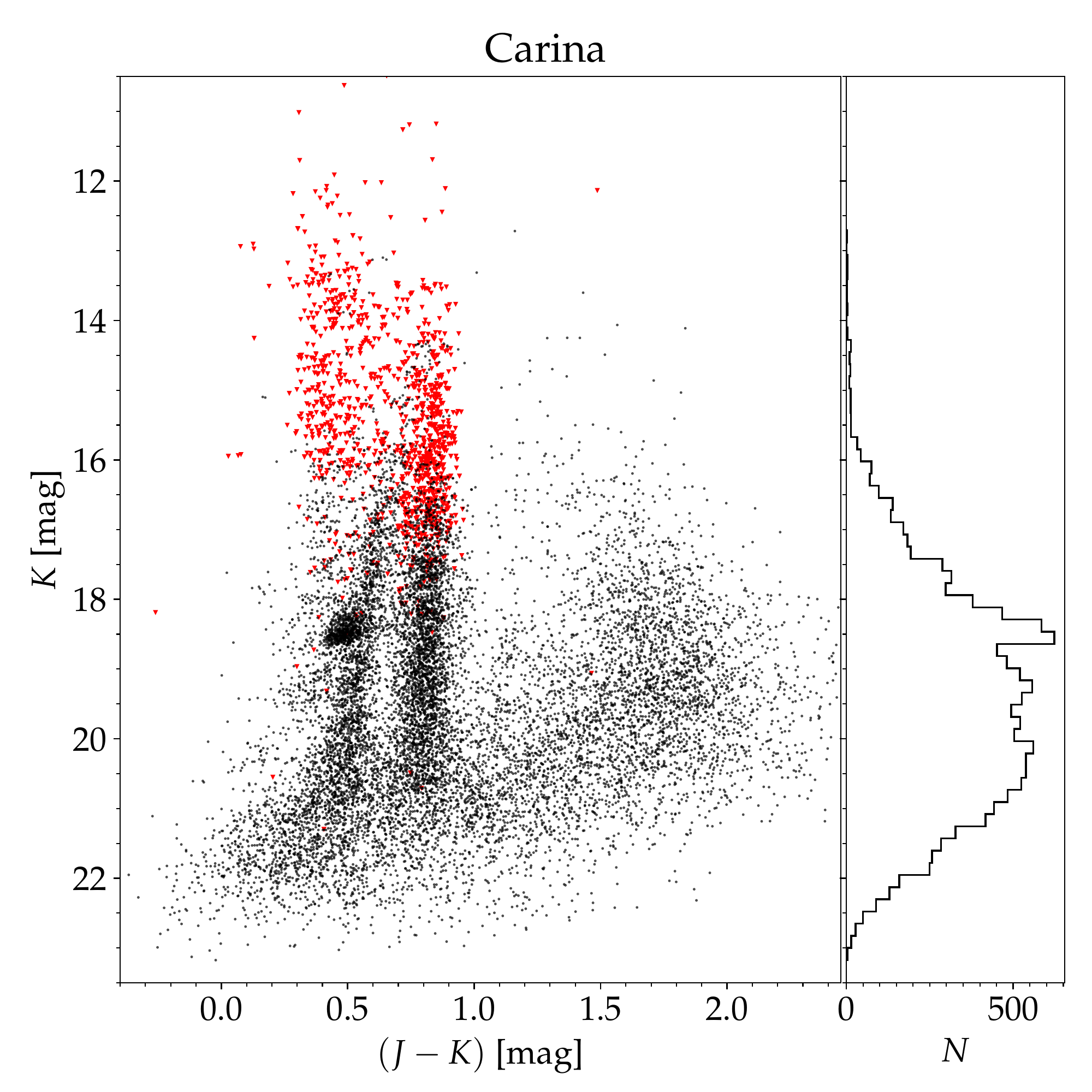}\\
    \includegraphics[width=0.79\hsize]{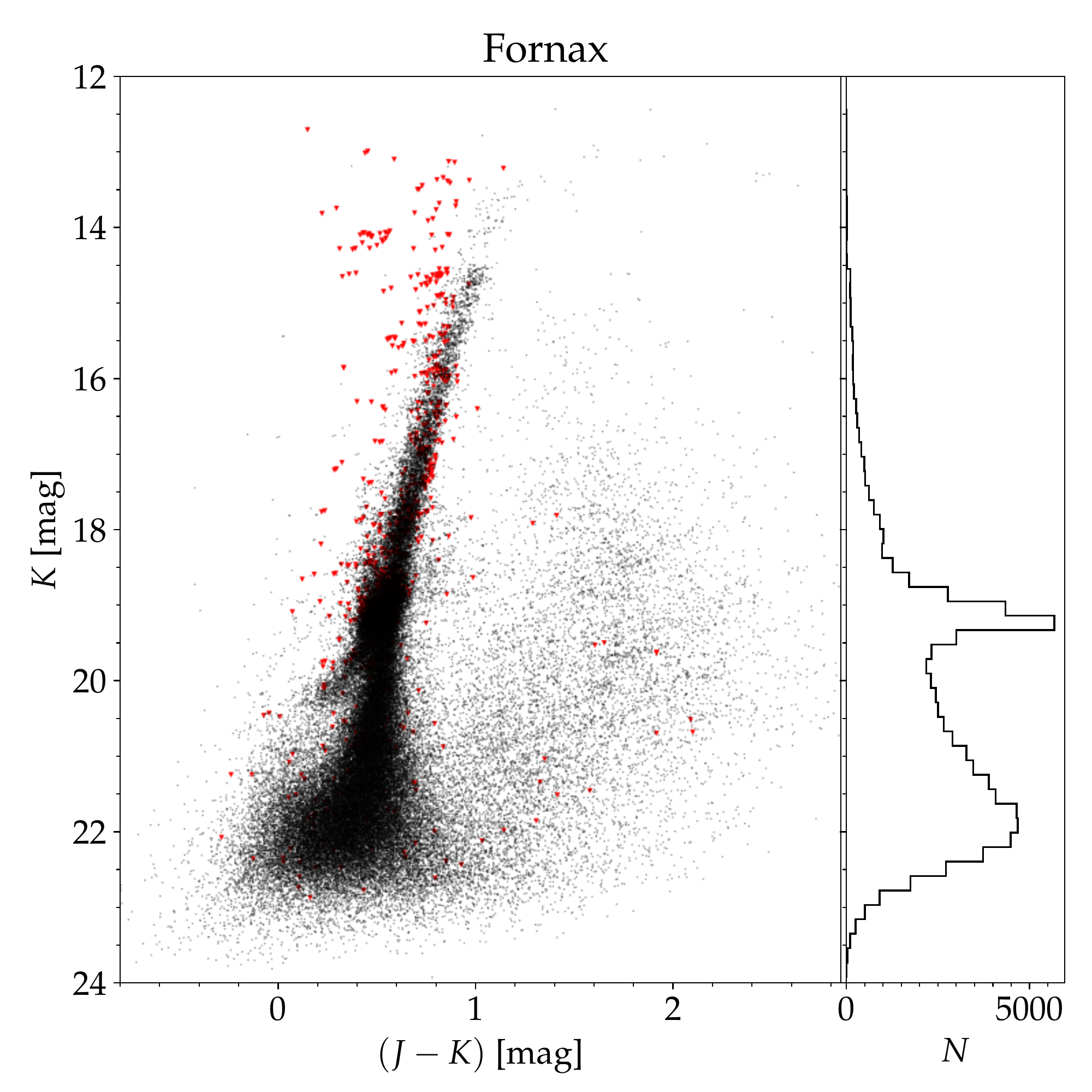}
    \caption{Color-magnitude diagrams of Sculptor, Carina and Fornax dwarf galaxies, obtained from a combination of the data of all observed fields. Note modest contamination with foreground stars (marked as red triangles) in Fornax and Sculptor, while considerable in Carina.}
    \label{fig:cmd}
\end{figure}

Our data are presented in Figure \ref{fig:cmd} in a form of $J-K$, $K$ color-magnitude diagrams (CMDs), with non-members marked as red triangles. The abundance of non-members in the CMD of Carina was expected due to Carina's galactic latitude of only $-22^\circ$, which places it relatively close to the Milky Way's plane. Moreover, the concentration of red triangles in two vertical pillars at $(J-K)\approx$ 0.4 and 0.8\,mag, which denote Galactic disk dwarf stars and halo giant stars, respectively, reassures that these objects indeed belong to the Milky Way.

\begin{splitdeluxetable*}{cccccccccccBcrrrrccc}
\tablecaption{Carina database. Consecutive columns present: identification number of a star, sky coordinates valid for epoch J2015.5, individual $J$- and $K_\mathrm{s}$-band measurements with their uncertainties, ellipticities, timestamps of exposures, probability of an object being a foreground star, pixel-wise location of stars in the reference images, the name of instrument, and the reference image.}\label{tab:db}
\tablehead{
\colhead{star} &\colhead{RA} & \colhead{Dec} & \colhead{mag\_J} & \colhead{err\_J} & \colhead{mag\_K} & \colhead{err\_K} & \colhead{ell\_J} & \colhead{ell\_K} & \colhead{JD\_J} & \colhead{JD\_K} & \colhead{frg} & \colhead{x\_J} & \colhead{y\_J} & \colhead{x\_K} & \colhead{y\_K} & \colhead{instr.} & \colhead{refimg\_J} & \colhead{refimg\_K} \\
\colhead{id} & \colhead{[deg]} & \colhead{[deg]} & \colhead{[mag]} & \colhead{[mag]} & \colhead{[mag]} & \colhead{[mag]} & & & \colhead{[d]} & \colhead{[d]} & & \colhead{[pix]} & \colhead{[pix]} & \colhead{[pix]} & \colhead{[pix]} & &
}
\colnumbers
\startdata
1	&	100.368437	&	--51.041293	&	13.707	&	0.003	&	13.330	&	0.003	&	0.099	&	0.025	&	2454787.7737084	&	2454787.78808168	&	0	&	320.062	&	1548.911	&	319.981	&	1539.760	&	HAWK-I	&	Car-F01\_081117\_J\_C1	&	Car-F01\_081117\_K\_C1	\\
2	&	100.350777	&	--51.066875	&	13.983	&	0.003	&	13.569	&	0.003	&	0.036	&	0.042	&	2454787.7737084	&	2454787.78808168	&	0	&	697.682	&	685.469	&	697.592	&	676.500	&	HAWK-I	&	Car-F01\_081117\_J\_C1	&	Car-F01\_081117\_K\_C1	\\
3	&	100.330486	&	--51.025024	&	14.194	&	0.003	&	13.677	&	0.003	&	0.035	&	0.041	&	2454787.7737084	&	2454787.78808168	&	0	&	1125.097	&	2100.87	&	1124.898	&	2091.719	&	HAWK-I	&	Car-F01\_081117\_J\_C1	&	Car-F01\_081117\_K\_C1	\\
4	&	100.379376	&	--51.065876	&	14.211	&	0.003	&	13.819	&	0.003	&	0.033	&	0.032	&	2454787.7737084	&	2454787.78808168	&	0	&	90.249	&	717.425	&	90.285	&	708.407	&	HAWK-I	&	Car-F01\_081117\_J\_C1	&	Car-F01\_081117\_K\_C1	\\
5	&	100.305289	&	--51.076543	&	14.248	&	0.003	&	13.892	&	0.003	&	0.013	&	0.035	&	2454787.7737084	&	2454787.78808168	&	0	&	1664.381	&	361.100	&	1664.166	&	352.150	&	HAWK-I	&	Car-F01\_081117\_J\_C1	&	Car-F01\_081117\_K\_C1	\\
6	&	100.351225	&	--51.056840	&	14.324	&	0.003	&	13.314	&	0.003	&	0.019	&	0.024	&	2454787.7737084	&	2454787.78808168	&	NaN	&	687.239	&	1024.572	&	687.111	&	1015.517	&	HAWK-I	&	Car-F01\_081117\_J\_C1	&	Car-F01\_081117\_K\_C1	\\
7	&	100.323382	&	--51.078197	&	14.944	&	0.003	&	14.574	&	0.003	&	0.040	&	0.020	&	2454787.7737084	&	2454787.78808168	&	0	&	1280.375	&	304.343	&	1280.213	&	295.401	&	HAWK-I	&	Car-F01\_081117\_J\_C1	&	Car-F01\_081117\_K\_C1	\\
8	&	100.348157	&	--51.060762	&	15.033	&	0.002	&	14.687	&	0.002	&	0.023	&	0.017	&	2454787.7737084	&	2454787.78808168	&	0	&	752.771	&	892.183	&	752.666	&	883.143	&	HAWK-I	&	Car-F01\_081117\_J\_C1	&	Car-F01\_081117\_K\_C1	\\
9	&	100.335028	&	--51.035396	&	15.191	&	0.002	&	14.463	&	0.002	&	0.027	&	0.044	&	2454787.7737084	&	2454787.78808168	&	0	&	1029.453	&	1750.144	&	1029.268	&	1740.981	&	HAWK-I	&	Car-F01\_081117\_J\_C1	&	Car-F01\_081117\_K\_C1	\\
10	&	100.359271	&	--51.031051	&	15.327	&	0.004	&	14.494	&	0.003	&	0.024	&	0.021	&	2454787.7737084	&	2454787.78808168	&	0	&	513.843	&	1895.571	&	513.711	&	1886.379	&	HAWK-I	&	Car-F01\_081117\_J\_C1	&	Car-F01\_081117\_K\_C1\\
\enddata
\tablecomments{A complete table is available on Zenodo \citep{karczmarek20db} and Araucaria website. A portion is shown here for guidance regarding its form and content. Analogous tables are available online for Fornax and Sculptor dwarf galaxies.}
\end{splitdeluxetable*}

\subsection{Comparison with near-infrared surveys}
\label{sec:comparisonsurveys}
Due to considerate overlap of targets, we were able to compare our databases with the content of three NIR surveys: the VISTA Hemisphere Survey \citep[VHS;][]{VHS13,VHSdb19}, the Two Micron All-Sky Survey \citep[2MASS;][]{TWOMASSdb03,TWOMASS06}, and the Deep Near-Infrared Survey of the Southern Sky \citep[DENIS;][]{DENIS98,DENISdb05}. The comparison covers the number of common stars, the photometric zero point, the depth of photometry, and the mode of observations (single/multi-epoch).

The only overlapping area between VHS and our observations is in the Carina galaxy. We cross-matched 9\,903 objects, and compared their magnitudes (after the transformation of our photometry to the VISTA photometric system). The photometric zero points are consistent within uncertainties, as was already shown in Table \ref{tab:zp_check}. Limiting magnitudes at a level of 5$\sigma$ for VHS are $J=22.0$\,mag  and $K=18.1$\,mag, and our limiting magnitudes at a level of 5$\sigma$ for Carina are $J=22.9$\,mag and $K=21.6$\,mag. Both VHS and our database consist of single-epoch observations.

We cross-matched our stars with DENIS targets and found 1\,242 common stars in Carina and 268 in Sculptor; there was no overlap with Fornax. We refrained from the analysis of the photometric zero-point for DENIS due to the same reason as mentioned by \citet{carpenter01}, i.e. the spread in the linear fit between 2MASS and DENIS data was significantly greater than expected based on random noise, and thus the transformation between the systems is not satisfactory. In order to evaluate the accuracy of the photometric zero point, both UKIRT and DENIS photometric systems would have to be transformed to a common photometric system, e.g. 2MASS, but such a transformation for DENIS may yield unreliable results. DENIS limiting magnitudes are $J=16.5$\,mag and $K=14.0$\,mag, and the observations were by design single-epoch, except the images taken under bad weather conditions, which were repeated. Our observations are single-epoch as well, with a few exceptions, but our photometry is considerably deeper.

The cross-match between 2MASS and our targets resulted in 1\,316 common stars in Carina, 3\,682 in Fornax and 143 in Sculptor. The zero points of 2MASS and our photometry are consistent within uncertainties (Table \ref{tab:zp_check}). Single-epoch observations in the 2MASS catalog are limited at a level of 10$\sigma$ to magnitudes $J=15.8$\,mag and $K=14.3$\,mag, making our photometry about 4-6\,mag deeper.

Lastly, we find no overlap between our fields and the UKIRT Hemisphere Survey Data Release 1 \citep[UHS DR1;][]{UHS18} nor the UKIRT Infrared Deep Sky Survey Data Release 11 \citep[UKIDSS DR11;][]{UKIDSS07}.

\subsection{Availability}
The databases of Carina, Fornax and Sculptor, together with supplementary data (completeness tables and plots, FITS images) are available online on Zenodo, DOI: \url{10.5281/zenodo.4396294} \citep{karczmarek20db}, and on the Araucaria website\footnote{\url{araucaria.camk.edu.pl/index.php/nir-photometric-maps}}.

\section{Conclusions}
\label{sec:conclusions}
We present deep NIR photometric maps of Carina, Fornax and Sculptor galaxies, and make them publicly available in a form of three digital tabulated databases. Our databases are comparable in depth to other NIR surveys, but cover areas of high scientific interest, that have been lacking quality photometry. To elevate the usability of our databases we provide additional information, like ellipticity of objects, completeness plots and tables, and FITS files of scientific quality, that can double as finding charts. We make our databases available through the Araucaria website and online servers. This paper is the first in a series of deep photometric maps of Local Group galaxies. The forthcoming papers will cover other galaxies, that have distances determined by the Araucaria Project.

\acknowledgments
We thank anonymous referee for thorough reading of the manuscript and a concise and informative report that helped to improve this paper. We acknowledge the contribution of present and former collaborators and members of the Araucaria Project, whose collective effort made this database possible.

The research leading to these results has received funding from the European Research Council (ERC) under the European Unions Horizon 2020 research and innovation program (grant agreement No 695099). We also acknowledge support from  Polish Ministry of Science and Higher Education, grant IdP II 2015 0002 64;  National Science Center, Poland, grant MAESTRO UMO-2017/26/A/ST9/00446; Millenium Institute of Astrophysics (MAS) of the Iniciativa Cientifica Milenio del Ministerio de Economia, Fomento y Turismo de Chile, project IC120009; BASAL Centro de Astrofisica y Tecnologias Afines (CATA), grant AFP-170002; FONDECYT POSTDOCTORADO, grant 3170703.

We are grateful to ESO committee for awarded time, within programs: 068.D-0352(A), 074.D-0318(B), 074.D-0505(B), 077.D-0423(B), 082.D-0123(A), 082.D-0123(B), 086.D-0078(B), 088.D-0401(B), 092.D-0295(B), and to ESO staff for carrying out the observations in service mode. 

This work has made use of 'Aladin sky atlas' developed at CDS, Strasbourg Observatory, France \citep{aladin00} and 
data from the European Space Agency (ESA) mission {\it Gaia} (\url{https://www.cosmos.esa.int/gaia}), processed by the {\it Gaia} Data Processing and Analysis Consortium (DPAC,
\url{https://www.cosmos.esa.int/web/gaia/dpac/consortium}). Funding for the DPAC has been provided by national institutions, in particular the institutions participating in the {\it Gaia} Multilateral Agreement, and general-purpose open-access repository Zenodo \citep{zenodo}.

\vspace{5mm}
\facilities{NTT(SOFI), VLT(HAWK-I, ISAAC)}

\software{astropy \citep{astropy13}, SExtractor \citep{sextractor96}, DAOPHOT/ALLSTAR \citep{stetson87}, IRAF \citep{iraf_tody86}, astroalign \citep{astroalign20}, Aladin \citep{aladin00}}

\newpage
\appendix
\section{Night log of observations}
\label{app:nightlog}
{\startlongtable
\begin{deluxetable*}{lccccccc}
\tablecaption{Journal of observations.}\label{tab:nightlog}
\tablehead{
\colhead{Field} & \colhead{RA (J2000)} & 
\colhead{Dec (J2000)} & \colhead{Date} & 
\colhead{Seeing} & \colhead{Filter} & 
\colhead{Exp. time} & \colhead{Instrument} \\
\colhead{name\tablenotemark{a}} & \colhead{[hh:mm:ss.s]} & \colhead{[dd:mm:ss]} & \colhead{} & 
\colhead{["]} & \colhead{} & \colhead{[min]} 
} 
\startdata
    \hline\multicolumn{8}{c}{Carina}\\
    \hline
    Car-F23 & 06:41:26.9 & --50:58:33 & 2001-11-09 & 1.12 & $J$ & 4  & ISAAC \\
    Car-F23 & 06:41:26.9 & --50:58:33 & 2001-11-09 & 1.19 & $K$ & 24 & ISAAC \\
    Car-F22 & 06:40:59.8 & --50:59:11 & 2001-11-09 & 0.95 & $J$ & 4  & ISAAC \\
    Car-F22 & 06:40:59.8 & --50:59:11 & 2001-11-09 & 0.84 & $K$ & 24 & ISAAC \\
    Car-F24 & 06:42:09.3 & --50:55:09 & 2001-11-09 & 0.89 & $J$ & 2  & ISAAC \\
    Car-F24 & 06:42:09.3 & --50:55:09 & 2001-11-09 & 1.04 & $K$ & 24 & ISAAC \\
    Car-F16 & 06:41:14.3 & --51:00:33 & 2001-11-10 & 1.08 & $J$ & 6  & ISAAC \\
    Car-F16 & 06:41:14.3 & --51:00:33 & 2001-11-10 & 1.05 & $K$ & 24 & ISAAC \\
    Car-F25 & 06:41:57.8 & --50:54:34 & 2001-11-10 & 1.56 & $J$ & 2  & ISAAC \\
    Car-F25 & 06:41:57.8 & --50:54:34 & 2001-11-10 & 1.87 & $K$ & 24 & ISAAC \\
    Car-F01 & 06:41:06.2 & --51:01:10 & 2008-11-17 & 1.48 & $J$ & 15 & HAWK-I\\
    Car-F01 & 06:41:06.2 & --51:01:10 & 2008-11-17 & 1.48 & $K$ & 31 & HAWK-I\\
    Car-F02 & 06:41:32.3 & --51:07:23 & 2008-11-17 & 1.32 & $J$ & 15 & HAWK-I\\
    Car-F02 & 06:41:32.3 & --51:07:23 & 2008-11-17 & 1.54 & $K$ & 31 & HAWK-I\\
    Car-F06 & 06:40:25.2 & --51:08:52 & 2008-11-17 & 1.39 & $J$ & 15 & HAWK-I\\
    Car-F06 & 06:40:25.2 & --51:08:52 & 2008-11-17 & 1.30 & $K$ & 31 & HAWK-I\\
    Car-F03 & 06:42:14.6 & --51:01:59 & 2008-11-20 & 0.90 & $J$ & 15 & HAWK-I\\
    Car-F03 & 06:42:14.6 & --51:01:59 & 2008-11-20 & 0.79 & $K$ & 31 & HAWK-I\\
    Car-F04 & 06:42:06.6 & --51:07:53 & 2008-11-20 & 0.72 & $J$ & 15 & HAWK-I\\
    Car-F04 & 06:42:06.6 & --51:07:53 & 2008-11-20 & 0.72 & $K$ & 31 & HAWK-I\\
    Car-F05 & 06:41:58.4 & --50:53:22 & 2008-11-20 & 0.91 & $J$ & 15 & HAWK-I\\
    Car-F05 & 06:41:58.4 & --50:53:22 & 2008-11-20 & 0.83 & $K$ & 31 & HAWK-I\\
    Car-F02 & 06:41:32.3 & --51:07:23 & 2008-12-05 & 1.49 & $J$ & 15 & HAWK-I\\
    Car-F02 & 06:41:32.3 & --51:07:23 & 2008-12-05 & 1.44 & $K$ & 31 & HAWK-I\\
    Car-F06 & 06:40:25.2 & --51:08:52 & 2008-12-05 & 1.16 & $J$ & 15 & HAWK-I\\
    Car-F06 & 06:40:25.2 & --51:08:52 & 2008-12-05 & 1.13 & $K$ & 31 & HAWK-I\\
    Car-F01 & 06:41:05.4 & --51:01:08 & 2013-11-27 & 0.93 & $J$ & 1.7 & SOFI\\
    Car-F01 & 06:41:05.4 & --51:01:08 & 2013-11-27 & 0.88 & $K$ & 1.9 & SOFI\\
    Car-F02 & 06:41:31.5 & --51:07:21 & 2013-11-27 & 1.00 & $J$ & 6.3 & SOFI\\
    Car-F02 & 06:41:31.5 & --51:07:21 & 2013-11-27 & 0.82 & $K$ & 4.2 & SOFI\\
    Car-F03 & 06:42:13.1 & --51:02:04 & 2013-11-27 & 0.98 & $J$ & 6.3 & SOFI\\
    Car-F03 & 06:42:13.1 & --51:02:04 & 2013-11-27 & 0.87 & $K$ & 4.2 & SOFI\\
    Car-F04 & 06:42:04.8 & --51:07:48 & 2013-11-27 & 0.93 & $J$ & 6.3 & SOFI\\ 
    Car-F04 & 06:42:04.8 & --51:07:48 & 2013-11-27 & 0.85 & $K$ & 4.2 & SOFI\\
    Car-F05 & 06:41:57.7 & --50:53:26 & 2013-11-27 & 0.91 & $J$ & 6.3 & SOFI\\ 
    Car-F05 & 06:41:57.7 & --50:53:26 & 2013-11-27 & 0.84 & $K$ & 4.2 & SOFI\\ 
    Car-F06 & 06:40:26.7 & --51:08:55 & 2013-11-27 & 0.96 & $J$ & 6.3 & SOFI \\
    Car-F06 & 06:40:26.7 & --51:08:55 & 2013-11-27 & 0.87 & $K$ & 4.2 & SOFI \\
    \hline\multicolumn{8}{c}{Fornax}\\
    \hline
    For-F12 & 02:40:30.8 & --34:25:13 & 2001-11-09 & 0.89 & $J$ & 4  & ISAAC \\
    For-F12 & 02:40:30.8 & --34:25:13 & 2001-11-09 & 0.81 & $K$ & 26 & ISAAC \\
    For-F13 & 02:40:08.7 & --34:11:28 & 2001-11-09 & 0.87 & $J$ & 4  & ISAAC \\ 
    For-F13 & 02:40:08.7 & --34:11:28 & 2001-11-09 & 0.51 & $K$ & 26 & ISAAC \\
    For-F11 & 02:39:32.9 & --34:34:24 & 2001-11-10 & 1.39 & $J$ & 13.3 & ISAAC \\ 
    For-F11 & 02:39:32.9 & --34:34:24 & 2001-11-10 & 1.41 & $K$ & 26   & ISAAC \\
    For-F14 & 02:39:35.8 & --34:25:37 & 2001-11-10 & 1.08 & $J$ & 13.3 & ISAAC \\ 
    For-F14 & 02:39:35.8 & --34:25:37 & 2001-11-10 & 1.35 & $K$ & 26   & ISAAC \\
    For-F01 & 02:40:22.4 & --34:24:52 & 2008-10-07 & 1.03 & $J$ & 18 & HAWK-I\\
    For-F01 & 02:40:22.4 & --34:24:52 & 2008-10-07 & 0.70 & $K$ & 26 & HAWK-I\\
    For-F03 & 02:39:39.6 & --34:25:06 & 2008-10-08 & 0.80 & $J$ & 18 & HAWK-I\\
    For-F03 & 02:39:39.6 & --34:25:06 & 2008-10-08 & 0.66 & $K$ & 26 & HAWK-I\\
    For-F03d & 02:39:39.6 & --34:25:06 & 2008-10-08 & 0.66 & $K$ & 45 & HAWK-I\\
    For-F04 & 02:40:35.7 & --34:33:42 & 2008-10-09 & 0.72 & $J$ & 18 & HAWK-I\\
    For-F04 & 02:40:35.7 & --34:33:42 & 2008-10-09 & 0.83 & $K$ & 26 & HAWK-I\\
    For-F04d & 02:40:35.7 & --34:33:42 & 2008-10-09 & 0.83 & $K$ & 43 & HAWK-I\\
    For-F01 & 02:40:22.4 & --34:24:52 & 2008-10-09 & 0.76 & $J$ & 18 & HAWK-I\\
    For-F01 & 02:40:22.4 & --34:24:52 & 2008-10-09 & 0.70 & $K$ & 26 & HAWK-I\\
    For-F01d & 02:40:22.4 & --34:24:52 & 2008-10-09 & 0.70 & $K$ & 45 & HAWK-I\\
    For-F02 & 02:40:08.5 & --34:29:45 & 2008-10-23 & 1.46 & $J$ & 18 & HAWK-I\\
    For-F02 & 02:40:08.5 & --34:29:45 & 2008-10-23 & 1.46 & $K$ & 26 & HAWK-I\\
    For-F01 & 02:40:20.9 & --34:25:11 & 2013-11-27 & 1.01 & $J$ & 6.3 & SOFI\\
    For-F01 & 02:40:20.9 & --34:25:11 & 2013-11-27 & 0.86 & $K$ & 4.2 & SOFI\\
    For-F02 & 02:40:09.8 & --34:29:15 & 2013-11-27 & 1.00 & $J$ & 6.3 & SOFI\\
    For-F02 & 02:40:09.8 & --34:29:15 & 2013-11-27 & 0.83 & $K$ & 4.2 & SOFI\\
    For-F04 & 02:40:35.1 & --34:33:35 & 2013-11-27 & 0.86 & $J$ & 19 & SOFI\\
    For-F04 & 02:40:35.1 & --34:33:35 & 2013-11-27 & 0.89 & $K$ & 25 & SOFI\\
    \hline\multicolumn{8}{c}{Sculptor}\\
    \hline
    Scu-F01d & 01:00:01.6 & --33:42:36 & 2004-12-31 & 0.92 & $J$ & 21 & SOFI \\
    Scu-F01d & 01:00:01.6 & --33:42:36 & 2004-12-31 & 0.91 & $K$ & 44  & SOFI \\
    Scu-F02d & 01:00:27.5 & --33:42:51 & 2004-12-31 & 0.85 & $J$ & 21 & SOFI \\
    Scu-F02d & 01:00:27.5 & --33:42:51 & 2004-12-31 & 0.88 & $K$ & 44  & SOFI \\
    \enddata
\tablenotetext{a}{Each name consists of a galaxy name followed by a field number with a suffix 'd' in case of deeper photometry.}
\end{deluxetable*}
}

\bibliography{bib}{}
\bibliographystyle{aasjournal}
\end{document}